\documentclass[letterpaper, 10 pt, conference]{ieeeconf}  

\IEEEoverridecommandlockouts                              

\usepackage[utf8]{inputenc}
\usepackage{include_packages} 
\usepackage[font=footnotesize,skip=5pt]{caption}
\usepackage[english]{babel}

\title{Avoiding Unintended Consequences:\\ How Incentives Aid Information Provisioning in Bayesian Congestion Games}
\author{{Bryce L. Ferguson, Philip N. Brown, and Jason R. Marden}
\thanks{This research was supported by ONR Grant \#N00014-20-1-2359, AFOSR Grant \#FA95550-20-1-0054, and NSF Grant \#ECCS-2013779}
\thanks{B. L. Ferguson (corresponding author) and J. R. Marden are with the Department of Electrical and Computer Engineering, University of California, Santa Barbara, CA, {\texttt{\{blferguson,jrmarden\}@ece.ucsb.edu}}.}
\thanks{P.N. Brown is with the Department of Computer Science, University of Colorado at Colorado Springs, {\texttt{\{philip.brown\}@uccs.edu}}.}
}

\begin{document}
\maketitle

\begin{abstract}
%
%
When users lack specific knowledge of various system parameters, their uncertainty may lead them to make undesirable deviations in their decision making.
To alleviate this, an informed system operator may elect to \emph{signal} information to uninformed users with the hope of persuading them to take more preferable actions.
In this work, we study public and truthful signalling mechanisms in the context of Bayesian congestion games on parallel networks.
We provide bounds on the possible benefit a signalling policy can provide with and without the concurrent use of monetary incentives.
We find that though revealing information can reduce system cost in some settings, it can also be detrimental and cause worse performance than not signalling at all.
However, by utilizing both signalling and incentive mechanisms, the system operator can guarantee that revealing information does not worsen performance while offering similar opportunities for improvement.
These findings emerge from the closed form bounds we derive on the benefit a signalling policy can provide.
We provide a numerical example which illustrates the phenomenon that
revealing more information can degrade performance when incentives are not used and improves performance when incentives are used.
\end{abstract}

\section{Introduction}
In many of today's large-scale cyber-physical or socio-technical systems, operating conditions are greatly affected by the decisions of the systems users, e.g., power grid demand~\cite{Palensky2011}, transit traffic~\cite{Pigou1920}, and supply-chain commerce or marketing~\cite{Esmaeili2009,Oreagba2021}.
Though users often make decisions in their own self interest, the system behavior that emerges need not be optimal~\cite{Hardin1968}.
This inefficiency can be further exacerbated by users' uncertainty over the state of the system~\cite{liu2016effects}, e.g., uncertainty on current electricity prices, traffic rates, or product quality.
By increasing their sensing and communication technologies, the operators of these systems gain the opportunity to learn the unknown system parameters and may choose to fully or partially share this information with their uninformed users.


The emergence of these new sensing and communication technologies opens the door to new methods for improving system performance.
One such method is that of information signalling by a well informed central authority~\cite{Kamenica2011,Goldstein2016,bergemann2019information,sezer2021social}.
By partially revealing their information about system parameters to uninformed users, the signaller allows the system users the opportunity to form new beliefs about their environment.
If the signaller reveals this information strategically, they may alter user behavior in such a way that the overall system performance is improved; for example, Google/Apple maps can reveal the travel times of certain routes to guide driver decision making in a way that can alter aggregate driving patterns and improve performance for the user population~\cite{Lindsey2014}.
One may initially think that all information should be shared with the users; however, this need not be optimal and could further degrade system performance.
The main focus of this work is understanding when and how a system operator should inform users about the system state and how much they can improve performance by doing so.

We study the principles of information signalling in the context of \emph{Bayesian congestion games}, where a group of users must route themselves through a congestable network while the exact congestion characteristics of each path are unknown.
The users possess a prior belief over system parameters that determine each paths congestion rate, for example, the belief there was an accident on a road or the chance weather has affected driving conditions.
The system operator can determine the values of these parameters and decide how to strategically reveal this information to their system users.
This model has been used to study the how signalling policies should be designed~\cite{Castiglioni2020,Zhu2021,wu2019information}, what behavior is likely to emerge~\cite{Wu2018}, and the associated performance of specific signalling structures~\cite{Tavafoghi2018,Lindsey2014}.
The results are typically limited to computational methods for finding signalling policies or a binary classification of whether information can or cannot help and frequently study simple, two path settings.
Additionally, for ease of analysis, much of the work in this area often assumes the signals are private (sent to individual users) and can be deceitful.
To better understand the true efficacy of revealing information, we consider public and fully truthful signals.
The main results of this work provide closed form bounds on the capabilities information signalling can provide in a parallel Bayesian congestion game with arbitrary polynomial latency functions.

Signalling mechanisms are becoming a topic of increasing research in their ability to influence user behavior, however, this is not the only influencing mechanism at a system operator's disposal.
Incentive mechanisms, where users are assessed monetary penalties or rewards based on their actions, have long been studied as an effective means of coordinating system behavior~\cite{Fleischer2004,Ferguson2019,Ferguson2021e,Cole2003,Paccagnan2021I}.
The interplay between incentives and signalling is an emerging area of study, and has up until now been limited to studying mechanisms where users must pay to acquire information \cite{Palma2012,Egorov2020,Heydaribeni2021}.
To the best of our knowledge, no current work has studied how monetary incentives and information signalling can be used concurrently to improve system performance.

In this work, we provide bounds on the possible benefit a signalling policy can provide with and without the concurrent use of monetary incentives.
When the system designer uses only information signalling, \cref{thm:untoll} shows that though signalling has the capability to improve system performance, there exist scenarios where revealing information can create worse system behavior.
Conversely, when the system operator can concurrently use signalling and incentive mechanisms, we show that they can guarantee that adding information does not worsen system performance while still offering similar capabilities for improvement.
Interestingly, we find that the possible benefit and detriment of information signalling increases with the complexity of the model of the unknown parameters and the amount of randomness.

We bolster these conclusions with a numerical example in \cref{sec:sim}, in which we find that without incentives, providing more information to the users makes system performance worse; however, when utilizing incentives, providing more information to the users improves system performance.


\section{Model}
\subsection{Congestion Games}
In this work, we consider the framework of network congestion games.
Consider a parallel, directed graph as a set $E$ of $n$ edges that connect a single source $s$ to terminal $t$ and can be traversed by a mass of traffic modeled by a continuum of infinitesimal agents.
Let $f_e \geq 0$ denote the mass of traffic utilizing edge $e \in E$, and let $f = \{f_e\}_{e \in E}$ denote a \emph{flow} in the network where $r = \sum_{e \in E} f_e$ is the rate of traffic or demand in the network\footnote{Without loss of generality, we can assume $r=1$, the proof of this is in Remark 1 in the appendix.}.

When a larger number of users traverse the same path, the congestion on that path increases.
To characterize this, each edge is endowed with a \emph{latency function} $\ell_e(f_e) = \sum_{d \in \mathcal{D}}\alpha_{e,d} (f_e)^d$, where $\alpha_{e,d} \geq 0$ for all $e \in E$ and $d \in \mathcal{D}$.
The latency function is a positive, increasing, convex polynomial where $\mathcal{D} =(d_1,\ldots,d_k)$ and $d_i \in \mathbb{Z}_{\geq 0}$ expresses the possible degrees, e.g., $\mathcal{D}=\{0,1\}$ represents affine congestion rates~\cite{Ferguson2019}, $\mathcal{D} = \{0,4\}$ can represent the well-known Bureau of Public Roads (BPR) latency functions, commonly used to model the congestion characteristics of physical roads~\cite{Farokhi2015b,Singh2002}, and $\mathcal{D} = \{0,\ldots,D\}$ can represent any positive, convex, increasing polynomial up to degree $D$~\cite{Paccagnan2021I}.
A congestion game of this form can thus be described by the tuple $G = (E,r,\{\alpha_{e,d}\}_{e \in E, d \in \mathcal{D}})$.
Let $\alpha \in \mathbb{R}_{\geq 0}^{|E|\cdot|\mathcal{D}|}$ be a vector containing the coefficients of each polynomial term in each latency function.
Notice that the size and values in $\alpha$ encode the size of the network and each edge's latency function; as such, we will use $\alpha$ as a variable that encodes all relevant information about the network structure throughout.

The overall congestion in the system can be measured by the \emph{total latency} 
\begin{equation}\label{eq:total_lat_det}
\Lat(f;\alpha) = \sum_{e \in E} f_e \ell_e(f_e),
\end{equation}
where each $\ell_e(\cdot)$ is inherently determined by $\alpha$.
We denote an optimal flow by $\opt{f} \in \argmin_{f} \Lat(f;\alpha)$.
However, the emergent traffic flow need not be determined by a central planner.
When users can choose their own path through the network, let $x \in [0,r]$ denote the index of an infinitesimal agent who uses an edge $e_x$.
When agents are left to choose routes that minimize their own observed travel time, i.e., each agent possesses a cost function $J_x(e;f) = \ell_e(f_e)$, then a plausible behavior that can emerge in the system is that of a \emph{Nash flow} $\Nash{f}$, which satisfies
\begin{equation}\label{eq:Nash_def}
J_x(e_x;f) \leq J_x(e^\prime;f),\quad \forall e^\prime \in E,~x \in [0,r].
\end{equation}
These system states are those where no user has an incentive to change their action and need not be optimal~\cite{Roughgarden2002};
additionally, the total latency in any Nash flow in a game of this form is the same~\cite{Mas-Colell1984}.

\subsection{Information Signalling}
To see what opportunities a system designer has in reducing the system's total latency in a Nash flow, we consider the role of information signalling.
Consider that the vector $\alpha$ is a random variable with distribution $\mu_0(x) = \mathbb{P}[\alpha=x]$ and support $A$, where
the system operator has knowledge of this parameter's realization, while the system users do not.
If the users reach a flow $f$, we extend \eqref{eq:total_lat_det} to be the \emph{expected total latency} over a distribution $\mu$, 
$$\Lat(f;\mu) = \underset{\alpha \sim \mu}{\mathbb{E}}\left[ \sum_{e \in E} f_e \ell_e(f_e) \right].$$
Because $\ell_e(\cdot)$ is determined by $\alpha$, it is a random variable.
We highlight two important possible realizations of the random variable $\alpha$ that will be used throughout: $\bota{\alpha} \in \mathbb{R}_{\geq 0}^{|E|\cdot |\mathcal{D}|}$ such that $\bota{\alpha}_{e,d} = \inf \{{\rm supp}(\alpha_{e,d})\}$ (where $\rm supp(\cdot)$ denotes the support of $\alpha$) in which each parameter takes its lowest value, and $\topa{\alpha} \in \mathbb{R}_{\geq 0}^{|E|\cdot |\mathcal{D}|}$ such that $\topa{\alpha}_{e,d} = \sup \{{\rm supp}(\alpha_{e,d})\}$ in which each parameter takes its largest value. Note that $\bota{\alpha}$ and $\topa{\alpha}$ need not be in the support of $\alpha$, but rather represent the corners of the smallest box that contains the support of $\alpha$ that are closest and furthest from the origin respectively.

The random variable $\alpha$ can effectively be used to model many forms of uncertainty, e.g., accidents or hazards in a traffic network or uncertain demand in network routing (see Remark 1).
With this purpose of $\alpha$ in mind, we include the following assumption:
\begin{assumption}\label{ass:pos}
In a Bayesian Congestion Game $G$ with prior $\mu_0$, $0,1 \in \mathcal{D}$ is always satisfied, and, for each edge $e \in E$, $\bota{\alpha}_{e,0},\bota{\alpha}_{e,1} >0$.
\end{assumption}
This assumption prevents cases where traffic can be routed with zero delay and has zero affect on congestion. 

As a method to coordinate system behavior and induce more desirable system states, the system designer may utilize a \emph{signalling policy} $\pi$, in which they reveal only a space in which the realization lies.
The signalling policy can be parameterized as $\pi = \{\pi_1,\ldots,\pi_m\}$, where $\pi_i \subset A$, $\pi_i \neq \emptyset$, $\cup_{i =1}^m \pi_i = A$, and $\pi_i \cap \pi_j = \emptyset$ for all $i,j \in \{1,\ldots,m\}$.
The signalling policy $\pi$ thus forms a partition of the support of $\alpha$, where agents are told which subset $\pi_i$ the realization $\alpha$ is restricted to, allowing them to update their beliefs.
These signals are assumed to be \emph{public and truthful}, and can be used to alter the users' belief over the system state.
Let $\mu_i(x) = \mathbb{P}[\alpha = x | \alpha \in \pi_i]$ denote the posterior belief users form when they receive $\pi_i$, and let $p_i = \int_{x \in \pi_i} \mu_0(x)dx$ denote the probability signal $\pi_i$ is sent.

Under a signalling policy $\pi$, agents may change their chosen path based on which signal they receive.
Let $\sflow = \{f(\pi_i)\}_{i=1}^m$ denote the tuple containing a flow that occurs at the reception of each signal, and let $\sigma_x = \{\sigma_x(\pi_i) \in E\}_{i=1}^m$ denote the edge user $x \in [0,r]$ selects after receiving each signal.
When each agent adopts a strategy based on the information system's signals, the system designer now cares about the expected total latency, expressed as
\begin{equation}
\Lat(\sflow;\mu_0,\pi) = \sum_{i=1}^m p_i  \Lat(f(\pi_i);\mu_i).
\end{equation}
An agent's cost will now be their expected travel time,
$$J_x(\sigma_x;\sflow,\mu_0,\pi) = \sum_{i=1}^m p_i  \underset{\alpha \sim \mu_i}{\mathbb{E}}\left[ \ell_{\sigma_x(\pi_i)}\left(f_{\sigma_x(\pi_i)}(\pi_i)\right) \right] .$$
We can now define a \emph{Bayes-Nash Equilibrium} as a tuple $(\BNash{\sflow},\BNash{\sigma})$ as a set of strategies where no agent elects to unilaterally change, i.e.,
\begin{multline}\label{eq:def_BNf}
J_x(\BNash{\sigma}_x;\BNash{\sflow},\mu_0,\pi) \leq J_x(\sigma^\prime;\BNash{\sflow},\pi,\mu_0),\\~ \forall \sigma^\prime \in E^m,~x \in [0,r].
\end{multline}

Our main focus in this work is understanding what opportunities a system designer has in lowering the expected total latency by way of information provisioning, i.e., comparing $\Lat(\BNash{\sflow};\mu,\pi)$ with $\Lat(\BNash{\sflow};\mu,\emptyset)$ (where the use of $\emptyset$ where a $\pi$ was expected denotes the case where no information is shared with users).
To quantify this improvement in system performance, we define the \emph{benefit to system cost} as our performance metric
\begin{equation}\label{eq:gain_def}
\Gain(\pi;\mu) = \Lat(\BNash{\sflow};\mu,\emptyset) - \Lat(\BNash{\sflow};\mu,\pi),
\end{equation}
which measures reduction in system cost from utilizing a signal policy $\pi$.
The system operator's objective is to institute a signalling structure that reduces the system cost or, equivalently, has a positive benefit.
The system operator need also avoid cases where the signalling policy increases the system cost or has a negative benefit to performance.

\subsection{Illustrative Example}
To illustrate the model, consider the Bayesian congestion game shown in \cref{fig:network_ex}.
In this problem instance, the network has two edges $E = \{1,2\}$, with latency functions $\ell_1(f_1) = \alpha_{1,0}$ and $\ell_2(f_2)=\alpha_{2,1}f_2$.
The latency coefficients $\alpha_{1,0},\alpha_{2,1}$ are unknown to the users and instead they possess a prior belief $\mu_0$ that is distributed over their support $A = [0,1]^2$.

Though the users do not observe the realizations of $\alpha_{1,0},\alpha_{2,1}$, they can receive some information from the system operator.
Under the signalling policy $\pi = \{\pi_1,\pi_2,\pi_3\}$, if the realization $\alpha \in \pi_i$ then the system operator sends a signal and the users update their belief to the Bayesian posterior $\mu_i$ with this knowledge.
For example if $\alpha_{1,0}=0.2$ and $\alpha_{2,1}=0.6$, then the system designer will inform users that $\alpha \in \pi_2$ and the users will update their beliefs to $\mu_2(x) = \mathbb{P}[x=\alpha|\alpha \in \pi_2]$.
Given a signalling policy $\pi$, the expected total latency of the Bayes-Nash flow is computed over $\alpha \in A$, where the flow changes based on which region $\alpha$ is realized.

{\begin{figure}[t!]
\vspace{1mm}
    \centering
    \begin{subfigure}[t!]{0.235\textwidth}
		\includegraphics[width=\textwidth]{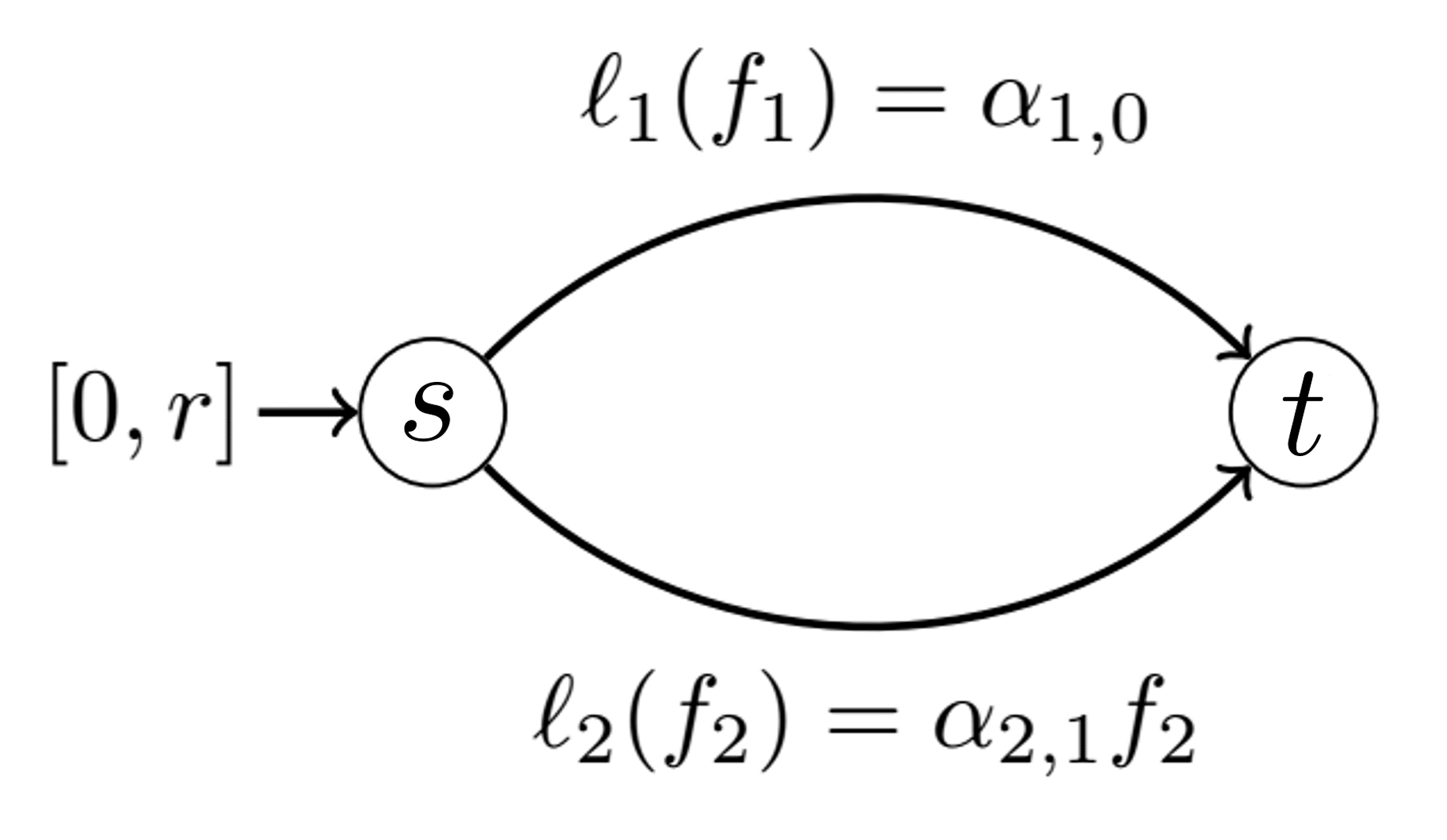}
    \end{subfigure}
    \begin{subfigure}[t!]{0.235\textwidth}
    	\includegraphics[width=\textwidth]{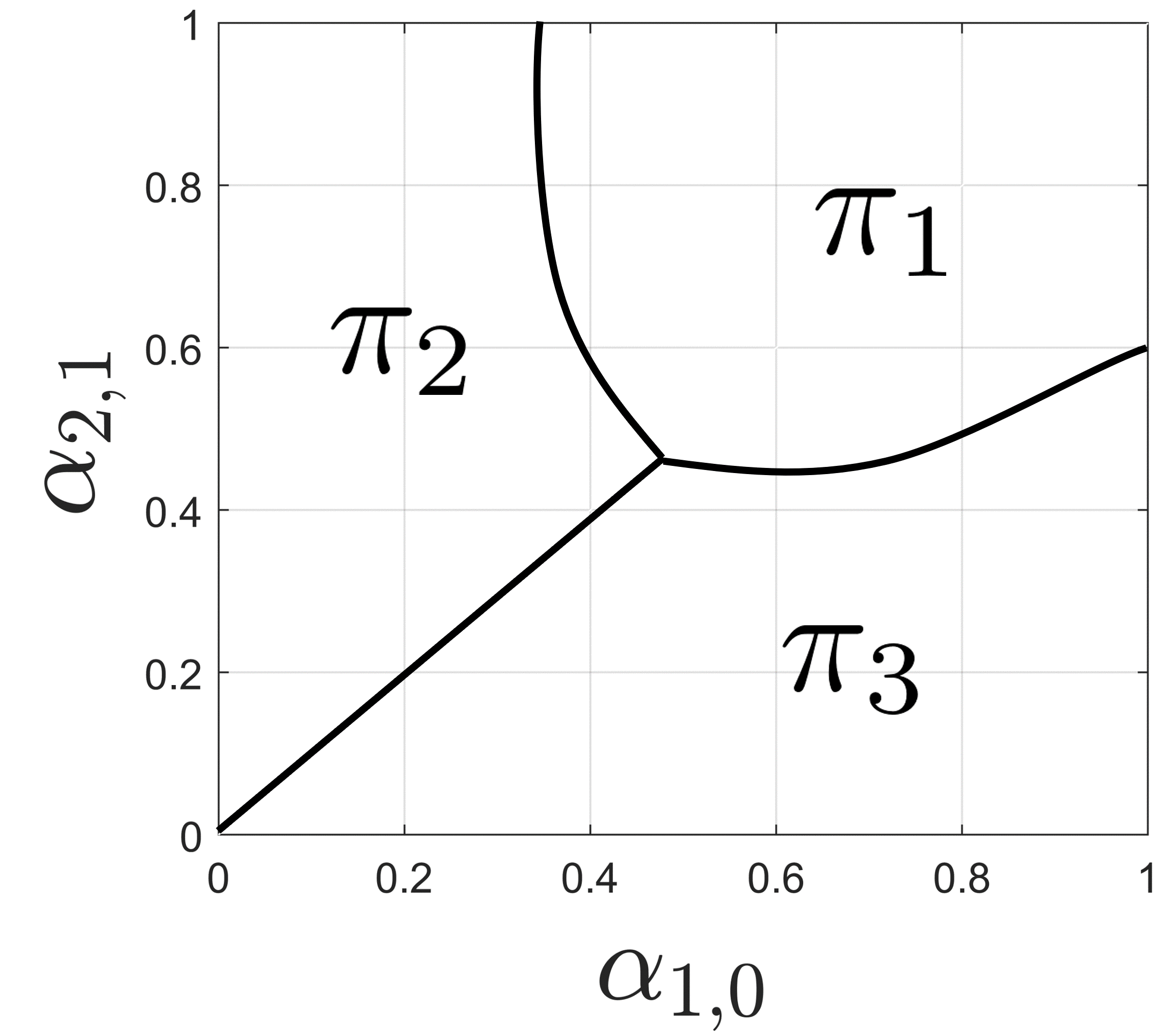}
    \end{subfigure}
    \caption{ {Two-link, parallel, Bayesian congestion game. One edge possesses a linear latency function, the other a constant latency function. The coefficients of each of these latency functions $\alpha_{1,0},\alpha_{2,1}$ are unknown but distributed with prior $\mu_0$ over $A=[0,1]^2$. A signalling policy $\pi = \{\pi_1,\pi_2,\pi_3\}$ partitions $A$ to map realizations to signals.}}
    \label{fig:network_ex}
    \vspace{-5mm}
\end{figure}
}

\subsection{Summary of Contributions}
The main contributions of this work come in characterizing the capabilities a system designer has in improving the system cost by revealing information about the system state to its users.
We do this by deriving bounds on the benefit a signalling policy can provide.
In \cref{sec:untoll}, we derive these bounds in the setting where an information provider can implement a public and truthful signalling policy;
In \cref{thm:untoll}, we find that there exists significant opportunities for the system designer to provide a positive benefit and improve performance via signalling, however, there is also the possibility that providing information has a negative benefit to system performance and can actually increase system cost, making it difficult for a system designer to guarantee improvement.

In \cref{sec:toll}, we add to the model by considering the case where a system designer need not only use an information signal, but may also design incentives they can levy on the users.
In this setting, we find in \cref{thm:toll} that when utilizing well designed tolls, the system designer has similar capabilities in reducing congestion but can now guarantee the signalling policy will never increase system cost, i.e., the benefit is always positive.
These results show that implementing incentive and signalling mechanisms concurrently can help a system designer avoid worsening system performance while maintaining their capabilities to improve it.

Finally, in \cref{sec:sim}, we provide a numerical simulation of the benefit of signalling.
When using a specific from of signalling policy that forms a uniform grid over the support; \cref{fig:benefit} shows the benefit when more information is revealed.
We see in this example that without incentives, there is an increasingly negative benefit to revealing information, however with incentives, the benefit is positive.

\section{Signalling Alone}\label{sec:untoll}

Consider a realization of a congestion game $G$ with latency function coefficients $\alpha$.
Let $\Nash{\Lat}(\alpha)$ denote the total latency in a Nash flow in this network (recall that this value is unique for each $\alpha$, ~\cite{Mas-Colell1984}).
When a system designer seeks to improve system performance via a public and truthful information system, \cref{thm:untoll} provides bounds on the benefit a signalling policy can provide.
Note that any absent proofs appear in the appendix.

\begin{theorem}\label{thm:untoll}
Consider the class of parallel Bayesian congestion games with polynomial latency functions whose degrees come from the set $\mathcal{D}$. For any distribution over the latency coefficients $\mu_0$ and any signalling policy $\pi$, the benefit in the expected total latency of a Bayes-Nash flow from signalling satisfies
\begin{equation}\label{eq:untoll_bound}
-\Theta\lVert\mathbb{E}[\alpha]-\bota{\alpha}\rVert_2 \leq \Gain(\pi;\mu_0) \leq \Theta\lVert\mathbb{E}[\alpha]-\bota{\alpha}\rVert_2,
\end{equation}
where $\Theta := {|\mathcal{D} | + \frac{\rho^+ - \rho^-_0}{2\rho^-_1}\left( |E| + |\mathcal{D}|-1 \right)}$, $\rho^-_0 = \min_{e \in E} \bota{\alpha}_{e,0}$, $\rho_1^- = \min_{e \in E}\bota{\alpha}_{e,1}$, $\rho^+ = \max_{e \in E} \sum_{d \in \mathcal{D}} (d+1)\topa{\alpha}_{e,d}$, $\mathbb{E}[\alpha] = \int_{x \in A}x \mu_0(x) dx$, and $\bota{\alpha} \in \mathbb{R}_{\geq 0}^{|E|\cdot |\mathcal{D}|}$ such that $\bota{\alpha}_{e,d} = \inf \{{\rm supp}(\alpha_{e,d})\}$ for each $e \in E$, $d \in \mathcal{D}$.
Additionally, there exists a $\mu_0$ such that for any $\pi \neq \emptyset$,
\begin{equation}\label{eq:sup_untoll}
\Gain(\pi;\mu_0) = \sqrt{|\mathcal{D}|}\lVert\mathbb{E}[\alpha]-\bota{\alpha}\rVert_2.
\end{equation}
Further, if $d \in \mathcal{D}$ where $d > 0$, then there exists a $\mu_0$ such that for any $\pi \neq \emptyset$,
\begin{equation}\label{eq:inf_untoll}
\Gain(\pi;\mu_0) = -\lVert\mathbb{E}[\alpha]-\bota{\alpha}\rVert_2.
\end{equation}
\end{theorem}

\cref{thm:untoll} reveals the capabilities a signalling policy has in improving system performance.
It also reveals the reality that revealing information can make system performance worse.
The bounds for the benefit of a signalling policy depends on the number of terms considered in each latency function $|\mathcal{D}|$, the size of the network $|E|$, as well as the distance between the average system state and the edge of its support $\lVert\mathbb{E}[\alpha]-\bota{\alpha}\rVert_2$ and other terms that change with the support.
One can think that the number of latency terms $|\mathcal{D}|$ characterizes the complexity of the model of network congestion while $\lVert\mathbb{E}[\alpha]-\bota{\alpha}\rVert_2$ measures the amount of uncertainty about the system parameters.
Additionally, \eqref{eq:sup_untoll} and \eqref{eq:inf_untoll} shows that there exist situations where regardless of what signalling policy is chosen, revealing information can greatly benefit or hinder system performance. 

To prove \cref{thm:untoll}, we first propose three lemmas.
The proof of each appears in the appendix.
\cref{lem:BNash_as_Nash} shows how we can characterize Bayes-Nash flows by looking at the expected latency coefficients after receiving a signal. 

\begin{lemma}\label{lem:BNash_as_Nash}
With a prior $\mu_0$ and a signalling policy $\pi$, the Bayes-Nash flow $\BNash{\sflow}$ can be characterized by $\{\Nash{\overline{f}}(\pi_i)\}_{i=1}^m$, where $\Nash{\overline{f}}(\pi_i)$ is the Nash flow in the network $G$ with coefficients $\overline{\alpha}_i = \mathbb{E}[\alpha|\alpha \in \pi_i]$.
\end{lemma}

Next, as a step in proving the bounds of \cref{thm:untoll}, we show how we can calculate the expected total latency.

\begin{lemma}\label{lem:exp_in_exp}
For a given flow $f$, the expected total latency with distribution $\mu$ over the coefficients $\alpha$ is equal to the total latency in the network with the expected coefficients, i.e.,
$\Lat(f;\mu) = \Lat(f;{\mathbb{E}}_{\alpha \sim \mu}[\alpha]).$
\end{lemma}

Utilizing \cref{lem:BNash_as_Nash} and \cref{lem:exp_in_exp} together, we can see that the expected total latency in a Bayes-Nash flow is equal to the weighted average of the total latency in the expected network after receiving each signal.
This fact will be used in proving \cref{thm:untoll}.

Finally, in \cref{lem:D_bound}, we provide a fact about the function $\Nash{\Lat}(\alpha)$ which bounds the difference in total latency between any two realizations of edge latency coefficients.

\begin{lemma}\label{lem:D_bound}
Consider the class of parallel congestion games with polynomial latency functions with degrees coming from the set $\mathcal{D}$ with coefficients $\alpha \in A$. Let $a,b \in \mathbb{R}_{\geq 0}^{|E|\cdot |\mathcal{D}|}$ be two possible sets of coefficients for a congestion game with edge set $E$, then
\begin{equation}\label{eq:D_bound}
{|\mathcal{D} | + \frac{\rho^+ - \rho^-_0}{2\rho^-_1}\left( |E| + |\mathcal{D}|-1 \right)} \geq \frac{\Nash{\Lat}(a)-\Nash{\Lat}(b)}{||a-b||_2},
\end{equation}
where $\rho^-_0 = \min_{e \in E} \bota{\alpha}_{e,0}$, $\rho_1^- = \min_{e \in E}\bota{\alpha}_{e,1}$, and $\rho^+ = \max_{e \in E} \sum_{d \in \mathcal{D}} (d+1)\topa{\alpha}_{e,d}$.
\end{lemma}

\section{Signalling \& Incentives}\label{sec:toll}
\cref{thm:untoll} showed that revealing information has the possibility of increasing or decreasing system cost.
This section deals with this fact by studying the advantages of concurrently utilizing signals and monetary incentives to influence user behavior.
Monetary incentives are a well studied method of influencing user behavior~\cite{Fleischer2004,Ferguson2019,Ferguson2021e,Cole2003,Paccagnan2021I}, but,
to the authors' best knowledge, the use of information signalling and monetary incentives in tandem has yet to be studied in the context of traffic networks.

Consider a congestion game $G$; an incentive designer can apply an incentive $\tau_e \in \mathbb{R}$ to each edge $e \in E$ to change the cost experienced by users utilizing that edge, i.e.,
$$J_x(e;f) = \ell_e(f_e) + \tau_e.$$
To understand how incentives and signals interact, consider a \emph{signal-aware incentive mechanism} $T(\pi_i;\pi,\mu_0)$ that assigns tolls $\{\tau_e(\pi_i)\}_{i=1}^m$ dependent on the signal broadcast by the information provider.
A player $x \in [0,r]$ with the strategy $\sigma_x$ now observes an expected cost of
\begin{multline*}
J_x(\sigma_x;\sflow,\mu_0,\pi,T)\\ =\sum_{i=1}^m p_i  \underset{\alpha \sim \mu_i}{\mathbb{E}}\left[  \ell_{\sigma_x(\pi_i)}\left(f_{\sigma_x(\pi_i)}(\pi_i)\right) + \tau_{\sigma_x(\pi_i)}(\pi_i)\right].
\end{multline*}
The Bayes-Nash flow definition remains as shown in \eqref{eq:def_BNf}, but now with users' tolled cost.
We now seek to understand the effectiveness of jointly implementing a signalling policy $\pi$ and an incentive mechanism $T$.
As such, we extend the definition of \eqref{eq:gain_def} which quantifies the gain in system performance to include the effect of an incentive mechanism $T$, i.e.,
\begin{equation}\label{eq:gain_def_toll}
\Gain(\pi;\mu,T) = \Lat(\BNash{\sflow};\mu,\emptyset,T) - \Lat(\BNash{\sflow};\mu,\pi,T).
\end{equation}

\begin{figure*}[t!]
\vspace{2mm}
    \centering
    \includegraphics[width=0.85\textwidth]{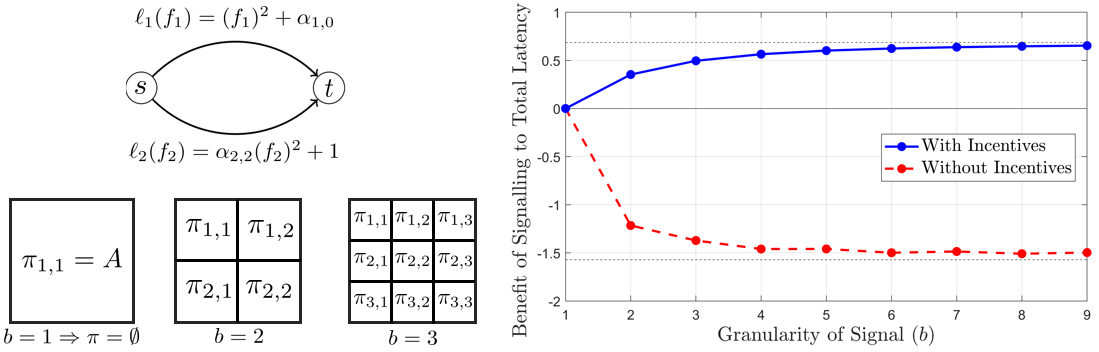}
    \caption{ {The benefit of revealing information with and without the concurrent use of incentives. $\pi$ is the uniform-grid signal structure where the support $A$ is partitioned into a grid with granularity $b$; as $b$ increases, more information is revealed to the users. At left, the benefit of using the uniform-grid signalling policy $\pi^b$ is shown for the setting described in \cref{sec:sim} with and without the concurrent use of the incentive mechanism $T^\star$. When incentives are used, revealing information provides a positive benefit and improves performance. With no incentives, the benefit becomes negative and revealing information worsens system cost.}}\label{fig:benefit}
    \vspace{-5mm}
\end{figure*}

In \cref{prop:opt_inc}, we first identify a signal-aware incentive mechanism that is optimal for a given signalling policy $\pi$.
\begin{proposition}\label{prop:opt_inc}
Let $\mu_0$ be a prior on the coefficients in a parallel congestion game with polynomial latency functions, and let $\pi$ be a signalling policy.
An optimal signal aware incentive mechanism $T^\star$ (i.e., minimizes $\Lat(\BNash{\sflow};\mu_0,\pi,T)$) assigns tolls as
\begin{equation}\label{eq:inc_def}
\Star{\tau}_e(\pi_i) = \sum_{d \in \mathcal{D}} d \underset{\alpha_{e,d} \sim \mu_i}{\mathbb{E}}[\alpha_{e,d}] (x_e)^{d},
\end{equation}
where $x \in \argmin_{f} \Lat(f;\mathbb{E}_{\alpha \sim \mu_i}[\alpha])$.
\end{proposition}
This incentive mechanism is reminiscent of a marginal cost toll in the deterministic setting, and possesses many of the same properties in terms of influencing Nash flow behavior to be optimal.
Consider a realization of a congestion game $G$ with latency function coefficients $\alpha$.
Let $\Star{\Lat}(\alpha)$ denote the total latency in the Nash flow in this network under the incentive mechanism $T^\star$.

\cref{thm:toll} provides bounds on the benefit a signalling policy can provide while also utilizing the signal-aware incentive mechanism $\Star{T}$.
We see that by concurrently utilizing incentives and signalling, the system designer can guarantee the benefit of signalling is non-negative.

\begin{theorem}\label{thm:toll}
Consider the class of parallel Bayesian congestion games with polynomial latency functions whose degrees come from the set $\mathcal{D}$. For any distribution over the latency coefficients $\mu_0$ and any signalling policy $\pi$, the decrease in the expected total latency of a Bayes-Nash flow from signalling satisfies
\begin{equation}\label{eq:toll_bound}
0 \leq \Gain(\pi;\mu_0,\Star{T}) \leq \Xi\lVert\mathbb{E}[\alpha]-\bota{\alpha}\rVert_2,
\end{equation}
where $\Xi := |\mathcal{D} | + \frac{\rho^+ - \rho^-_0}{4\rho^-_1}\left( |E| +\sum_{d \in \mathcal{D}\setminus \{0\}} (d+1)^d \right)$, $\rho^-_0 = \min_{e \in E} \bota{\alpha}_{e,0}$, $\rho_1^- = \min_{e \in E}\bota{\alpha}_{e,1}$, $\rho^+ = \max_{e \in E} \sum_{d \in \mathcal{D}} (d+1)\topa{\alpha}_{e,d}$,
 $\mathbb{E}[\alpha] = \int_{x \in A}x \mu_0(x) dx$, and $\bota{\alpha} \in \mathbb{R}_{\geq 0}^{|E|\cdot |\mathcal{D}|}$ such that $\bota{\alpha}_{e,d} = \inf \{{\rm supp}(\alpha_{e,d})\}$ for each $e \in E$, $d \in \mathcal{D}$.
\end{theorem}

%
%

Comparing the bounds on the benefit of a signalling policy with and without the use of incentives (i.e., \eqref{eq:untoll_bound} and \eqref{eq:toll_bound}), we see that incentives can make the use of signals more robust and guarantee they improve performance while allowing for similar opportunities to improve performance.
We further support this conclusion in \cref{sec:sim} by providing a numerical example and comparing the benefit of revealing information with and without incentives.

Before we prove \cref{thm:toll}, we state the following lemma that is similar to \cref{lem:D_bound} but applies to $\Star{\Lat}$.

\begin{lemma}\label{lem:D_bound_star}
Consider the class of parallel congestion games with polynomial latency functions with degrees coming from the set $\mathcal{D}$ with coefficients $\alpha \in A$. Let $a,b \in \mathbb{R}_{\geq 0}^{|E|\cdot |\mathcal{D}|}$ be two possible sets of coefficients for a congestion game with edge set $E$, then
\begin{equation}\label{eq:D_bound_star}
|\mathcal{D} | + \frac{\rho^+ - \rho^-_0}{4\rho^-_1}\left( |E| + \hspace{-3pt}\sum_{d \in \mathcal{D}\setminus \{0\}}\hspace{-2pt} (d+1)^d \right) \geq \frac{\Star{\Lat}(a)-\Star{\Lat}(b)}{||a-b||_2}.
\end{equation}
where $\rho^-_0 = \min_{e \in E} \bota{\alpha}_{e,0}$, $\rho_1^- = \min_{e \in E}\bota{\alpha}_{e,1}$, $\rho^+ = \max_{e \in E} \sum_{d \in \mathcal{D}} (d+1)\topa{\alpha}_{e,d}$.
\end{lemma}

\section{Numerical Example}\label{sec:sim}
To understand how the benefit of signalling changes as more information is revealed, we offer the following numerical example.
Consider a Bayesian congestion game with two edges, $\ell_1(f_1) = f_1^2+\alpha_{1,0}$ and $\ell_2(f_2) = \alpha_{2,2}f_2^2+1$, where $\alpha_{1,0}$ and $\alpha_{2,2}$ are parameters unknown to the user.
We consider that these two parameters are drawn from a truncated normal distribution, i.e.,
let
$$z \sim \mathcal{N}\left(\begin{bmatrix}30\\30\end{bmatrix},180\times\begin{bmatrix}2&1\\1&2\end{bmatrix}\right),$$
and define the prior over $\alpha_{1,0},\alpha_{2,2}$ as $\mu_0(x) = \mathbb{P}[x=z|z\in A]$ where $A = [0,60]^2$ is their support.

Now, we analyze the benefit of the \emph{uniform-grid signalling policy} with and without the concurrent use of the signal-aware incentive mechanism $\Star{T}$ as defined in \cref{prop:opt_inc}.
Let $b$ be an integer representing the granularity of the signalling mechanism, i.e., the number of times $A$ is partitioned along each dimension as shown in \cref{fig:benefit}, i.e.,
$$\pi_{i,j} =  \left[\frac{60}{b}(i-1),\frac{60}{b}i\right] \times \left[\frac{60}{b}(j-1),\frac{60}{b}j\right],$$
essentially forming a uniform grid over $A$.

In \cref{fig:benefit}, we plot the benefit for using the uniform-grid signalling policy with and without the concurrent use of the incentive mechanism $\Star{T}$, i.e., $\Gain(\pi;\mu_0)$ and $\Gain(\pi;\mu_0,\Star{T})$.
Observe that when no incentives are used, increasing the amount of information revealed to users (i.e., larger $b$) causes the benefit to become increasingly negative; meaning as more information is revealed, the signalling policy makes the system performance worse.
Conversely, while using incentive mechanism $\Star{T}$, as more information is added, the benefit becomes increasingly positive and revealing information now improves performance.

\section{Conclusion}
This work studies the capabilities of an information signalling system with and without the concurrent use of incentives.
We show that when signalling alone, revealing information has the possibility to increase system cost, however, with the concurrent use of incentives and signals, revealing information can only improve the system performance.
Future work will look into how to design optimal signalling policies while using incentives and will consider how to co-design information and signals when each mechanism has different amounts of available information (e.g., non-signal-dependent incentives or semi informed incentives).

\section{Acknowledgments}
We would like to thank the reviewers, particularly the anonymous reviewer \#10, for their comments which greatly improved the presentation and results of this paper.

\bibliographystyle{IEEEtran}
\bibliography{../../../../../../library}

\appendix
\begin{remark}\label{rem:demand}
Without loss of generality, we can assume a unit traffic rate, $r=1$, even when $r \sim \nu$ is a random variable.
\end{remark}
\noindent \emph{Proof:}
Consider a congestion game $G$ with demand $r$.
Define a mapping $Q(G,\gamma)$ that outputs a new congestion game $\hat{G}$ with latency functions $\hat{\ell}_e(x) = \sum_{d \in \mathcal{D}} \frac{\alpha_{e,d}}{\gamma^{d+1}}(f_e)^d$.
Let $f$ be a flow in $G$ with total traffic $r$.
Now, consider the flow $\gamma f = \{\gamma  f_e\}_{e \in E}$ in $\hat{G}$.
Each edge $e \in E$ will have latency 
$$\hat{\ell}_e(\gamma f_e) =\sum_{d \in \mathcal{D}} \frac{\alpha_{e,d}}{\gamma^{d+1}}(\gamma f_e)^d = \frac{1}{\gamma}\ell_e(f_e).$$
Notice that latency on each edge is scaled by $1/\gamma$ and the preference structure is preserved; therefore, if $f$ is a Nash flow in $G$ then $\gamma f$ is a Nash flow in $\hat{G}$.
Further, $$\Lat(\gamma f;\hat{G}) = \sum_{e \in E} \gamma f_e \hat{\ell}_e(\gamma f_e) =  \sum_{e \in E}  f_e {\ell}_e(f_e) = \Lat(f;G),$$
and the two networks will have the same total latency.

If $(\alpha,r) \sim \mu_0$, e.g., $\mu_0(x,y) = \mathbb{P}[\alpha=x,r=y]$, then consider that ${\alpha}^\prime \sim \hat{\mu}_0$, where $\hat{\mu}_0(z) = \sum_{x,y|Q(x,1/y)=z}\mu_0(x,y)$.
Now ${\alpha}^\prime$ has the same distribution over total latency.
\hfill\qed

\noindent \emph{Proof of \cref{lem:BNash_as_Nash}:}
Consider the prior $\mu_0$ on $\alpha$ and the signalling policy $\pi$.
If the signal $\pi_i$ is sent to users, they update their belief via Bayesian inference to
$\mu_i(x) = \frac{\mu_0(x)}{p_i}\indicator{x \in \pi_i}$.
In a flow $f$, user $x \in [0,r]$ experiences an expected cost of
\begin{align*}
J_x(e_x;f,\mu_i) &= \underset{\alpha \sim \mu_i}{\mathbb{E}}\left[ \sum_{d \in \mathcal{D}} \alpha_{e_x,d}(f_{e_x})^d \right]\\
&=\sum_{d \in \mathcal{D}} \mathbb{E}[\alpha_{e_x,d}|\alpha \in \pi_i](f_e)^d.
\end{align*}
Note that if $f$ were not a Nash flow in the parallel congestion game with coefficients $\mathbb{E}[\alpha|\alpha \in \pi_i]$, then \eqref{eq:Nash_def}, at least one user $x \in [0,r]$ would be able to deviate their strategy $\sigma_x(\pi_i)$ and experience lower cost.
Therefore, the only Bayes-Nash flows occur when $f(\pi_i)$ is a Nash flow with respect to $\mathbb{E}[\alpha|\alpha \in \pi_i]$ for all $i \in \{1,\ldots,m\}$.
Further, because the total latency in a Nash flow is unique, so too is the expected total latency in a Bayes-Nash flow.
\hfill\qed

\noindent \emph{Proof of \cref{lem:exp_in_exp}:}
Consider the distribution $\mu_0$ on $\alpha$, and let $f$ be a flow in the network.
The expected total latency
\begin{align*}
	\Lat(f;\mu) &= \underset{\alpha \sim \mu}{\mathbb{E}}\left[ \sum_{e \in E} f_e  \sum_{d \in \mathcal{D}} \alpha_{e,d}(f_{e})^d \right]\\
	&=  \sum_{e \in E} f_e  \sum_{d \in \mathcal{D}} \underset{\alpha \sim \mu}{\mathbb{E}}\left[\alpha_{e,d}\right](f_{e})^d \\
	&= \Lat(f;\underset{\alpha \sim \mu}{\mathbb{E}}\left[\alpha\right]),
\end{align*}
which follows from linearity of expected value.
\hfill\qed

\noindent \emph{Proof of \cref{lem:D_bound}}
We assume $r = 1$, which is without loss of generality from \cref{rem:demand}.
We note that $\Nash{\Lat}(\alpha)$ is continuous, but need not be continuously differentiable; as such we look for the largest gradient in the differentiable regions of the support.
Let $\Nash{f}(\alpha)$ be the Nash flow in the parallel congestion game with polynomial coefficients $\alpha$, i.e., $\Nash{\Lat}(\alpha) = \Lat(\alpha,\Nash{f}(\alpha)) = \sum_{e \in E} \sum_{d \in \mathcal{D}} \alpha_{e,d}(\Nash{f_e})^{d+1}$.
First, we seek to bound the partial derivative of $\Nash{\Lat}(\alpha)$ with respect to some parameter $\alpha_{e,d}$.
Clearly, $\frac{\partial}{\partial \alpha_{e,d}} \Nash{\Lat}(\alpha) \geq 0$ in parallel networks as no Braess's paradox type example can exist~\cite{MILCHTAICH2006321}.
To upper-bound this partial derivative, we will consider a case where by increasing $\alpha_{e,d}$ any mass of traffic that chooses to leave edge $e$ will all choose the edge $e^\prime$; in general, this may not occur with every change in $\alpha_{e,d}$, as users may disperse over multiple edges, however, if we consider that users do all move to the same edge, and we pick edge $e^\prime$ as the one that increases the total latency most rapidly, then the following upper-bound will hold.
With this in mind, it implies that $\frac{\partial}{\partial \alpha_{e,d}} \Nash{f_e} = -\frac{\partial}{\partial \alpha_{e,d}} \Nash{f}_{e^\prime}$, and that we can evaluate the partial derivative as
\begin{multline}\label{eq:partial_1}
\hspace{-11pt}\frac{\partial}{\partial \alpha_{e,d}} \Nash{\Lat}(\alpha) \hspace{-1pt} = \hspace{-1pt} (\Nash{f}_e)^{d+1} + \Bigg(\hspace{-3pt}\sum_{d^{\prime\prime} \in \mathcal{D}} \alpha_{e^\prime,d^{\prime\prime}}(d^{\prime\prime}+1)(\Nash{f}_{e^\prime})^{d^{\prime\prime}+1} \\ 
- \sum_{d^\prime \in \mathcal{D}} \alpha_{e,d^\prime}(d^\prime+1)(\Nash{f}_e)^{d^\prime}\Bigg) \frac{\partial}{\partial \alpha_{e,d}} \Nash{f}_{e^\prime}
\end{multline}
Now, we note that latency on edges $e$ and $e^\prime$ must be the same in a Nash flow, thus
\begin{align*}
\ell_e(\Nash{f}_e) &= \ell_{e^\prime}(\Nash{f}_{e^\prime})\\
\frac{\partial}{\partial \alpha_{e,d}}\ell_e(\Nash{f}_e) &= \frac{\partial}{\partial \alpha_{e,d}}\ell_{e^\prime}(\Nash{f}_{e^\prime}).
\end{align*}
Using this equality, and the fact that $\frac{\partial}{\partial \alpha_{e,d}} \Nash{f_e} = -\frac{\partial}{\partial \alpha_{e,d}} \Nash{f}_{e^\prime}$, we can evaluate the derivative and rearrange to get
\begin{equation}\label{eq:flow_grad_bound}
\frac{\partial}{\partial \alpha_{e,d}} \Nash{f}_{e^\prime} = \frac{(\Nash{f}_e)^d}{\ell_e^\prime(\Nash{f}_e) + \ell_{e^\prime}^\prime(\Nash{f}_{e^\prime})} \leq \frac{(\Nash{f}_e)^d}{2\rho^-_1},
\end{equation}
where $\rho^-_1 = \min_{e \in E} \bota{\alpha}_{e,1}$.
Substituting \eqref{eq:flow_grad_bound} into \eqref{eq:partial_1} gives us
\begin{equation}\label{eq:Nash_Lat_partial_bound}
\frac{\partial}{\partial \alpha_{e,d}} \Nash{\Lat}(\alpha) \leq (\Nash{f}_e)^{d+1} + \frac{\rho^+-\rho^-_0}{2\rho_1^-}(\Nash{f}_e)^d,
\end{equation}
where $\rho^-_0 = \min_{e \in E} \bota{\alpha}_{e,0}$ and $\rho^+ = \max_{e \in E} \sum_{d \in \mathcal{D}} (d+1)\topa{\alpha}_{e,d}$.

Now, the gradient of $\Nash{\Lat}(\alpha)$ must satisfy
\begin{align*}
\lVert \nabla \Nash{\Lat}(\alpha) \rVert_2 &\leq \sqrt{ \sum_{e \in E}\sum_{d \in \mathcal{D}} \left((\Nash{f}_e)^{d+1} \hspace{-1pt} + \hspace{-1pt} \frac{\rho^+-\rho^-_0}{2\rho_1^-}(\Nash{f}_e)^d \right)^2 }\\
&\leq \sqrt{ \left(\sum_{e \in E}\sum_{d \in \mathcal{D}} (\Nash{f}_e)^{d+1} \hspace{-1pt} + \hspace{-1pt} \frac{\rho^+-\rho^-_0}{2\rho_1^-}(\Nash{f}_e)^d \right)^2 }\\
&\leq \sum_{d \in \mathcal{D}} \left(\sum_{e \in E}\Nash{f}_e\right)^{d+1} + \frac{\rho^+ - \rho^-_0}{2\rho_1^-} \cdot \\
&~~~~~~~~~~~\left(\sum_{e \in E} (\Nash{f}_e)^0 \sum_{d \in \mathcal{D}\setminus \{0\}} \left( \sum_{e \in E} \Nash{f}_e \right)^d  \right)\\
&=  |\mathcal{D} | + \frac{\rho^+ - \rho^-_0}{2\rho^-_1}\left( |E| + |\mathcal{D}|-1 \right),
\end{align*}
where the first inequality holds from \eqref{eq:Nash_Lat_partial_bound}, the second and third hold from the super-additivity of convex monomials of positive terms, and the final equality holds from the assumption that $r=1$.

%

Finally, consider two sets of coefficients $a,b \in A$. Observe
\begin{align*}
&\frac{\Nash{\Lat}(a)-\Nash{\Lat}(b)}{\lVert a-b \rVert_2}\\
 &= \frac{1}{\lVert a-b \rVert_2} \int_{\lambda=0}^1(a-b)^T \nabla \Nash{\Lat}(\lambda a+ (1-\lambda)b) d\lambda\\
&\leq \frac{1}{\lVert a-b \rVert_2} \int_{\lambda=0}^1 \lVert a-b \rVert_2 \cdot \lVert\nabla \Nash{\Lat}(\lambda a+ (1-\lambda)b)\rVert_2 d\lambda\\
&\leq \int_{\lambda=0}^1 {|\mathcal{D} | + \frac{\rho^+ - \rho^-_0}{2\rho^-_1}\left( |E| + |\mathcal{D}|-1 \right)} ~d\lambda\\
& = {|\mathcal{D} | + \frac{\rho^+ - \rho^-_0}{2\rho^-_1}\left( |E| + |\mathcal{D}|-1 \right)}.
\end{align*}
Where the first inequality holds from Cauchy-Schwarz, and the second holds from our observation above on the norm of the gradient of $\Nash{\Lat}(\alpha)$.
\hfill\qed

\noindent \emph{Proof of \cref{lem:D_bound_star}:}
This proof follows very similarly to the proof of \cref{lem:D_bound}, but now, in an optimal flow $\Star{f}(\alpha)$, the latency on each edge is not equal, however, the marginal-cost on each edge is~\cite{Pigou1920}.
Let $\nu_e(f_e) = \sum_{d \in \mathcal{D}} (d+1)\alpha_{e,d} (f_e)^d$ be the marginal cost on edge $e$ with flow $f_e$.
Now, in the optimal flow $\Star{f}$ (which emerges from using the tolls $\Star{T}$),
\begin{align*}
\nu_e(\Star{f}_e) &= \nu_{e^\prime}(\Star{f}_{e^\prime})\\
\frac{\partial}{\partial \alpha_{e,d}}\nu_e(\Star{f}_e) &= \frac{\partial}{\partial \alpha_{e,d}}\nu_{e^\prime}(\Star{f}_{e^\prime}).
\end{align*}
Evaluating and rearranging these derivatives gives
\begin{equation}\label{eq:opt_flow_grad_bound}
\frac{\partial}{\partial \alpha_{e,d}} \Star{f}_{e^\prime} = \frac{(d+1)(\Star{f}_e)^d}{\nu_e^\prime(\Star{f}_e) + \nu_{e^\prime}^\prime(\Star{f}_{e^\prime})} \leq \frac{(d+1)(\Star{f}_e)^d}{4\rho^-_1}.
\end{equation}
With this, we can upper bound the partial derivative of $\Star{\Lat}$ as
\begin{equation}\label{eq:Opt_Lat_partial_bound}
\frac{\partial}{\partial \alpha_{e,d}} \Star{\Lat}(\alpha) \leq (\Star{f}_e)^{d+1} + \frac{\rho^+-\rho^-_0}{4\rho_1^-}(d+1)(\Star{f}_e)^d.
\end{equation}
Following the same steps as in the proof of \cref{lem:D_bound}, the gradient of $\Star{\Lat}$ must satisfy
\begin{align*}
\lVert \nabla \Star{\Lat}(\alpha) \rVert_2 \leq   |\mathcal{D} | + \frac{\rho^+ - \rho^-_0}{4\rho^-_1}\left( |E| +\sum_{d \in \mathcal{D}\setminus \{0\}} (d+1)^d \right).
\end{align*}
Finally, we can use this bound as in the proof of \cref{lem:D_bound} to complete the proof.

\noindent \emph{Proof of \cref{prop:opt_inc}:}
Consider that the users receive signal $\pi_i$ from signalling policy $\pi$ and prior $\mu_0$ (forming posterior $\mu_i$), and realize a flow of $f$.
From \ref{lem:exp_in_exp}, the expected total latency in this flow is equal to the total latency of this flow in the expected network, i.e., $\Lat(f;\mu_i) = \Lat(f;\overline{\alpha}_i)$, where $\overline{\alpha}_i = \mathbb{E}_{\alpha \sim \mu_i}[\alpha]$.
Thus, an optimal flow at the reception of signal $\pi_i$ is one that satisfies $\opt{f}(\pi_i) \in \argmin_f \Lat(f;\overline{\alpha}_i)$.

We now look for a incentives that will influence users such that $\opt{f}(\pi_i)$ becomes a Nash flow in a congestion game with latency coefficients $\overline{\alpha}_i$.
To do so, we note that $G$ with flow-varying incentive functions $\tau_e(f_e)$ is a potential game~\cite{Rosenthal1973} with potential function
$$\Phi(f;\alpha) = \sum_{e \in E} \int_{0}^{f_e} \ell_e(x) + \tau_e(x) dx.$$
As such, the flow in $\argmin_f \Phi(f;\alpha)$ is a Nash equilibrium.
For the polynomial latency functions considered in this work, let $\tau_e(x) = \sum_{d \in \mathcal{D}} d \alpha_{e,d} (x)^d$.
Now, the potential function becomes
\begin{align*}
\Phi(f;\alpha) &= \sum_{e \in E} \int_{0}^{f_e} \sum_{d \in \mathcal{D}} \alpha_{e,d}x^d + d \alpha_{e,d} x^d dx \\
&= \sum_{e \in E} \sum_{d \in \mathcal{D}} f_e\alpha_{e,d}(f_e)^d = \Lat(f;\alpha),
\end{align*}
and the Nash flow that minimizes $\Phi$ also minimizes $\Lat$; as such, $\opt{f}(\pi_i)$ becomes a Nash equilibrium in the game with the coefficients $\overline{\alpha}_i$.

Finally, notice that by selecting the fixed incentive $\Star{\tau}_e(\pi_i) = \tau_e(\opt{f}_e(\pi_i))$, the equilibrium conditions do not change and $\opt{f}(\pi_i)$ remains a Nash flow.
Nash flows retain the same uniqueness properties under fixed incentives, and thus assigning $\Star{\tau}(\pi_i)$ minimizes the expected total latency when $\pi_i$ is sent.
If this is done for each signal, the total latency with each signal will me minimal and so too will the overall expected total latency, making $T^\star$ an optimal incentive mechanism.
\hfill\qed

\end{document}